\documentclass [aps,prb,superscriptaddress,twocolumn,showpacs,
a4paper,unsortedaddress,amsmath,amssymb,byrevtex,floatfix]{revtex4}

\usepackage[latin1]{inputenc}
\usepackage{graphicx}
\usepackage{dcolumn}
\usepackage{bm}
\usepackage{hyperref}
\usepackage{times}

\begin{document}

\title{Universality in the low-voltage transport response of
molecular wires physisorbed onto graphene electrodes}

\author{V. M. Garc\'{\i}a-Su\'arez}
\affiliation{Departamento de F\'{\i}sica, Universidad de Oviedo,
33007 Oviedo, Spain}
\affiliation{Nanomaterials and Nanotechnology Research Center (CINN),
Spain}
\affiliation{Department of Physics, Lancaster University,
Lancaster, LA1 4YB, United Kingdom}

\author{R. Ferrad\'as}
\affiliation{Departamento de F\'{\i}sica, Universidad de Oviedo,
33007 Oviedo, Spain}
\affiliation{Nanomaterials and Nanotechnology Research Center (CINN),
Spain}

\author{D. Carrascal}
\affiliation{Departamento de F\'{\i}sica, Universidad de Oviedo,
33007 Oviedo, Spain}
\affiliation{Nanomaterials and Nanotechnology Research Center (CINN),
Spain}

\author{J. Ferrer}
\affiliation{Departamento de F\'{\i}sica, Universidad de Oviedo,
33007 Oviedo, Spain}
\affiliation{Nanomaterials and Nanotechnology Research Center (CINN),
Spain}
\affiliation{Department of Physics, Lancaster University,
Lancaster, LA1 4YB, United Kingdom}

\date{\today}

\begin{abstract}
We analyze the low-voltage transport response of large molecular wires
bridging graphene electrodes, where the molecules are physisorbed
onto the graphene sheets by planar anchor groups.
In our study, the sheets are pulled away to vary the gap length and the
relative atomic positions. The molecular wires are also translated
in directions parallel and perpendicular to the sheets.
We show that the energy position of the Breit-Wigner molecular resonances
is universal for a given molecule, in the sense that it is independent of
the details of the graphene edges, gaps lengths or of the molecule positions.
We discuss the need to converge carefully the $k-$sampling to provide
reasonable values of the conductance.
\end{abstract}


\maketitle
Single- or few-molecule electronics is considered a plausible
alternative technology to silicon, that could be deployed to enable the CMOS industry
to reach the atomic limit\cite{Tourbook,Fer09}. However, the few-particle nature
of the contacts in single-molecule devices leads to a lack of mechanical robustness.
It also leads to a large phase space of coupling configurations, each with a different
conductance. The spread in experimental values of the transport response is sorted
by the use of statistically averaged conductance histogram.
A large body of the research in this field has therefore concentrated on improving
the robustness of the devices by searching and testing new generations of contact
groups\cite{Kam09,Chen11,Ven06,Mar13}.

\begin{figure*}
\includegraphics[width=2\columnwidth]{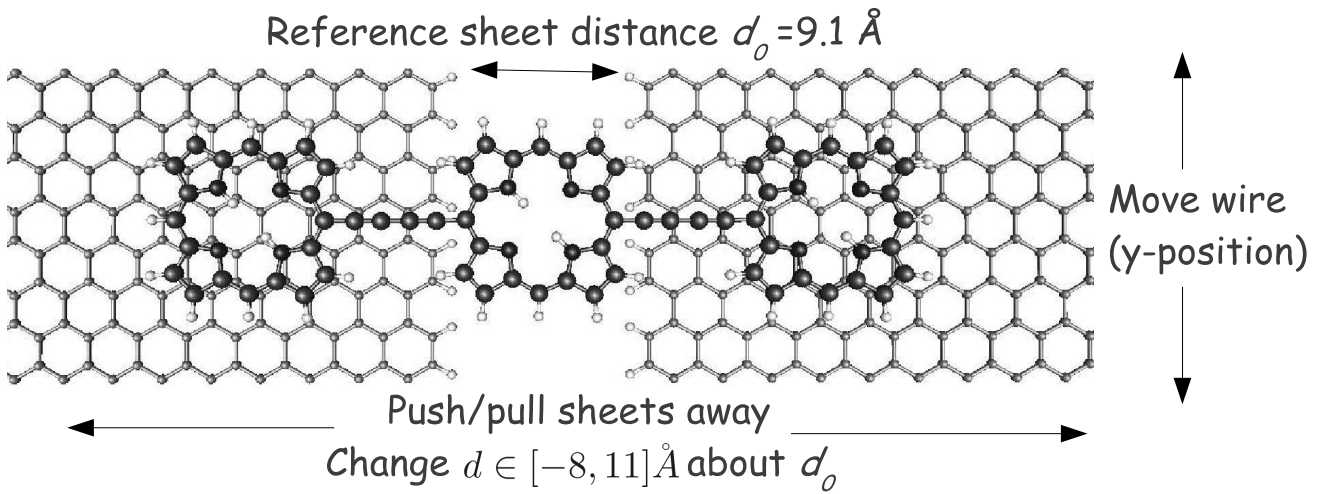}
\caption{\label{Junction}(Color online) Schematic plot of the
single-molecule junction discussed in this article.
Small light-grey (graphene)
and large dark-grey (molecule) spheres and small white spheres
represent carbon and hydrogen atoms, respectively. Medium-size
spheres around the center of each porphyrin ring represent
nitrogen atoms. The system has a total of 446 atoms.}
\end{figure*}

Graphene could replace noble-metals as the material for the electrodes to help
solving these key issues.
The use of graphene brings about several clear advantages. First, its lack of
atomic diffusivity makes each gap geometry stable over time.
Second, graphene's planar nature allows for the visualization of the junctions,
and possibly the control over the actual molecule positioning. Third, compared
to the more bulky metallic electrodes, the molecular junction can be placed closer
to a gate; graphene electrodes could lead to a reduced gate-field screening and
therefore to an enhanced gate coupling. Furthermore, the conductance of few-layer
graphene is largely gate-independent so that features of the contacted molecules
may not be masked by the electrodes response to the gate. Recently it was also
proven experimentally that stable and gateable molecular junctions can be formed
by depositing molecules on top of a graphene nanogap fabricated by electroburning,
where the molecules' anchor groups were probably contacting the edges \cite{Pri11}.

Graphene therefore deserves to be scrutinized as a possible material of choice
for single-molecule electrodes. But graphene brings about its own series of problems,
that must be overcome. We mention here its complex edges morphology, passivation and
oxidation states, that are not fully characterized yet. The ability to fabricate
nanogaps where gap length and edge morphology are controlled with almost atomic
accuracy is yet to be developed.

A number of  theoretical articles have proposed bridging graphene electrodes with molecules,
whose anchor groups would be chemisorbed to the electrodes' edges\cite{Kol07,Che08,Sah10,Sah11}.
However, we have shown in a previous article that the tribological nature of graphene edges
leads to a large variability in the transport response of these devices\cite{Car12}.
An alternative approach that avoids edge anchoring was advanced in Refs. (\onlinecite{Ber11,Pet12}), where
fullerene-based dumbbell wires where physisorbed onto graphene, or molecular
wires were physisorbed onto carbon nanotubes. In addition, by contacting the molecules deep inside
the sheets, trouble related to the size-mismatch between the physical gap and the
length of the molecule is avoided from the outset\cite{Ber11}. Notice that this strategy can
not be implemented with bulky noble-metal electrodes.

In this manuscript we explore the electronic properties and the low-voltage
electrical response of molecular wires physisorbed onto graphene leads.
Physisorption is a gentler attachment to graphene than chemisorption.
The distortion of the geometry and the electronic structure of both molecule and graphene's
contact area is expected to be only moderate and mostly related to screening
effects. Specifically, charge transfer between molecule and electrodes is strongly
suppressed so that the contact-induced dipoles are expected to be small.
In contrast to Ref. (\onlinecite{Ber11}), we analyze molecular wires with planar
anchor groups such as porphyrine or phthalocyanine  molecules. These groups are
linked by polyyne  chains to form dimer and trimer wires, an example of which is
shown in Fig. (\ref{Junction}). The rationale behind our choice is that increasing  the
contact surface area could enhance the mechanical stability of the junctions and provide an
effective conductance averaging. In addition, porphyrine wires show a small conductance decay
with wire length when contacted by gold electrodes\cite{Sed11}. Porphyrine
wires can also show interference and spin-filtering effects that make them
promising functional units\cite{Fer12}. As opposed to dumbbell
molecules, the central region in our wires is by geometrical constraints placed always
at the same height measured from the graphene layers, bringing a desirable degree of
isolation to the functional central unit and, therefore, of reproducibility,

Our main result is a remarkable universality in the energy position of the Breit-Wigner (BW)
molecular resonances, which is missing for junctions comprising noble-metal electrodes,
or for edge-contacted  graphene junctions. We find that the energy position of the
BW resonances  appearing in the zero-voltage transmission coefficient $T(E)$
does not shift when the molecules are displaced, or when the graphene sheets are
pulled away to vary the physical gap length.  We have observed this behavior not
only for the junction discussed in this manuscript but also for monomers,  dimers
or trimers made up of porphyrine or phthalocyanine planar groups physisorbed onto
graphene.
Our results suggest that it might be possible to attach universal energy positions
to the BW resonances of a given physisorbed molecular wire. These energy
positions could be controlled by a suitable gate in close proximity.
We have found however a large spread in the low-voltage conductance
$G=\mathrm{G}_0\,T(E=E_\mathrm{F})$
as a function of the molecule-electrodes relative positions, which we atribute
to the energy shift of Fano-like resonances around the Fermi level, driven by the
molecule-graphene relative motion.  We comment lastly on
the need to compute $T(E)$ using a large sampling in transverse
$k$-points. We find that $T(E\sim E_\mathrm{F})$ develops a cusp at the  Dirac point
$E_\mathrm{F}$, and that it changes by orders of magnitude from the $\Gamma-$point
estimate, as the accuracy in the $k-$summation is improved.
This contrast to the results found for gold junctions, where it was shown
that a few $k$-points are more than sufficient to find accurate transport results.

We discuss our results in terms of the junction shown in Fig. (\ref{Junction}), where
a trimer molecule made up of three porphyrin units is linked by butadiynes, e.g.:
atomic chains made of four carbon atoms. We find that the whole of this molecule is
placed roughly at the same height above the sheets ($\sim 3.2$ \AA), regardless
of its position. Hence, we discuss the changes in the transmission
coefficients and in the conductance when the molecule is displaced across the plane in
directions parallel and perpendicular to the edges, or when the graphene sheets are pulled
away or pushed towards each other. The sheets shown in Fig. (\ref{Junction}) are terminated
in an armchair configuration passivated by hydrogen atoms in a 1-1-1-sequence, which
is predicted to be the most stable reconstruction and passivation on this kind of
edge \cite{Was08,Kos08,Jia11}.  We stress however that our strategy to reduce
the conductance tribological effects brought about by the edges relies on
attaching the molecule anchor groups deep inside the graphene sheets.

We have used the Density Functional Theory (DFT) code SIESTA\cite{SIESTA},
which employs norm-conserving pseudopotentials to account for the core
electrons and linear combinations of atomic orbitals to construct
the valence states. In order to better describe physisorption,
we have used a recent implementation\cite{Rom09} of the van der Waals
functional of Dion et al\cite{Dio04}. We have computed the Hamiltonian
and overlap matrix elements, the potential and the density using
a real-space grid with fineness defined with an equivalent energy
cutoff of 200 Ry. We have relaxed quickly the coordinates to forces smaller
than 0.05 eV/\AA\ using a single-$\zeta$ basis set (SZ). We have taken then
these coordinates as seeds for further coordinate relaxation using a
single-$\zeta$ polarized (SZP). We have repeated the procedure now with a
double-$\zeta$ (DZ) and then with a double-$\zeta$ polarized (DZP). We have
first relaxed an isolated porphyrin ring on top of graphene and used
those coordinates as a starting point to construct the junction by
joining the carbon chains and the central porphyrin ring. Tests performed
with one of the junctions show that there are not very large differences
in the positions calculated with different basis sets. The
electronic structure is however very sensitive to the choice of the
basis: we have found large differences in the transmission and the
density of states when the basis set improved from SZ to SZP and to
DZ. There are not however many changes in moving from DZ to DZP,
which implies that the electronic structure is well converged with
DZ. We have used then this basis set to perform all calculations.
The junctions analyzed here are made periodic along the
direction perpendicular ($y$) to electronic transport ($z$).
Each electrode has
10 principal layers, where each principal layer has two columns of
carbon atoms. This is enough to accommodate the porphyrin molecules on
top of the sheets and ensure that the electronic structure converges to the bulk
electronic structure at the leftmost and rightmost layers.
We have also computed the physisorption energy for chemical detachment of this and
other molecules. We have found energy barriers of the order of 2 eV.

\begin{figure}
\includegraphics[width=0.95\columnwidth]{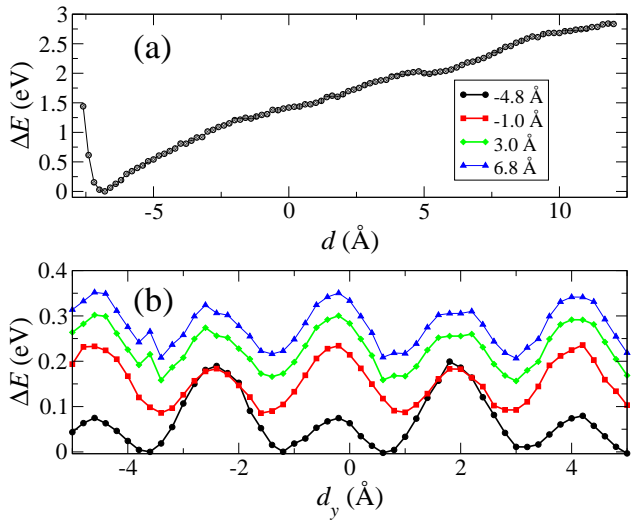}
\caption{\label{Fig2} (Color online)
Energy analysis of the Junction plotted in Fig. (\ref{Junction}):
Junction energy as a function of $d_z$ (a) and as a function of the lateral
displacement $d_y$ for several $d_z$ (b). In the last figure, the energy curves have
been displaced in energy to fit them in the same graph.}
\end{figure}

The graphene layers are pulled/pushed along the $z-$axis opposite
to each other, so that the physical gap length $d$ increases or decreases. We
have chosen a reference distance $d_0$ between layers
equal to 9.1 \AA. This corresponds to the
separation in which the central porphyrin covers entirely the gap.
We have varied the gap length in the range $d\in (-7.6, 11.2)$ \AA\, in steps
of 0.2 \AA. We note that with our basis set, atomic orbitals at opposing
electrodes cease to overlap for gap lengths $d \gtrsim -2$ \AA\,.
A plot in Fig. (\ref{Fig2}) (a) of the junction energy as a function of $d$ (e.g.: when
the electrodes are pulled away) shows a minimum at about $d \sim -6.8$ \AA.
The existence and position of the minimum reflects a trade-off between
two opposing factors: hydrogen-passivated edge repulsion versus molecule-graphene bonding.

We had computed previously the energy barriers for phthalocyanine monomers sliding
on top of an infinite graphene sheet. We found values $ \sim 40$ meV, which were
consistent with a previous calculation\cite{Lebedeva10}. Now, we have displaced sideways
the trimer molecule in Fig. (\ref{Junction}) for fixed $d$, so that it
drifts parallel to the physical gap.
Because the trimer has two porphyrine anchor groups bonded to the sheets as well as two
polyyne chains and a third anchor group interacting with the edges, we expect energy
barriers two or three times bigger that the result for the monomer. The junction energy,
plotted in Fig. (\ref{Fig2}) (b), shows a smooth  and periodic variation
of the energy, whose period reflects the underlying graphene lattice.
The energy barriers are of about 0.1 to 0.2 eV,
consistent with our previous analysis. We can make a rough
estimate of the molecule's drift length along the gap, $r \sim \sqrt{4\,D\,t}$, using an
Arrhenius equation for the diffusion constant
\begin{equation}
D=D_0\,e^{-E_\mathrm{B}/k_\mathrm{B}T}
\end{equation}
\noindent where we take $D_0\sim 10^{-4}$ cm$^2/$s and $E_B=0.2$ eV.
We find that the trimer would
stay fixed at its position for temperatures up to about 100 K, but it should
drift at room temperature. This quick analysis indicates that the trimer junction
would be stable if energy barriers of about 0.5 eV were achieved. Achieving such
high energy barriers may be done by attaching suitable side groups, or better tailoring
the anchor groups.

\begin{figure}
\includegraphics[width=\columnwidth]{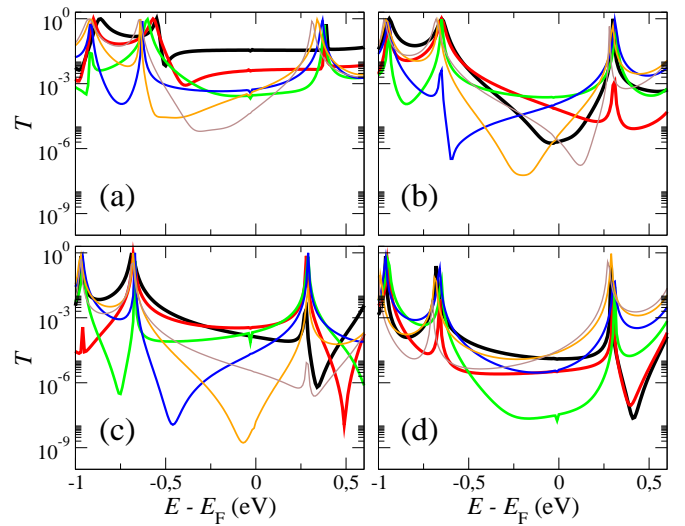}
\caption{\label{Fig3}(Color online) Evolution of the
$\Gamma-$point transmission coefficient $T(E)$ with increasing
gap length, $d$, which varies between -7.6 \AA\ and +10.8 \AA\ relative to $d_0$,
in steps of 0.8 \AA. Panels (a), (b) (c) and (d) correspond to
curves in the intervals [-7.6 \AA, -3.6 \AA], [-2.8 \AA, 1.2 \AA],
[2.0 \AA, 6.0 \AA] and [6.8 \AA, 10.8 \AA], respectively.
Decreasing width of the lines corresponds to increasing $d$.}
\end{figure}

The junction transport properties have been computed using the newly developed multi-scale
multi-terminal transport code GOLLUM \cite{gollum}. This program has a simple
interface which can easily read any tight-binding model or DFT Hamiltonian, which are
generically called Hamiltonian providers (HP). The program is much
faster than previous transport codes due to a new  implementation of zero- and
finite-voltage algorithms\cite{Gar12}.  We have used in the present calculation SIESTA as
the HP. We have performed $\Gamma-$point calculations to start with,
and have tested the convergence of the zero-voltage transmission coefficient $T(E)$
as a function of the number of  transverse $k$-points, as discussed below.

The $\Gamma-$point estimates of $T(E)$ for different gap lengths $d$ are plotted
in Fig. (\ref{Fig3}), at intervals of 0.8 \AA\cite{Dirtun}. The figure shows that
the position of the molecular BW resonances remains almost constant.
We stress that we have found the same
behavior for other junctions: the energy position of the BW resonances for a given
graphene-based junction does not depend on the molecule position relative to the physical
gap, provided that the bonding mechanism is by physisorption.
This universality could have been anticipated: physisorption carries no
charge transfer between the molecule and the sheets, nor associated dipole
moments. This is so because both sides of each contact are mostly made up of
the same chemical species: carbon. Also, the $\pi-\pi$ hybridization between
molecular orbitals and the electrode states is weaker than for the
bonds present in most noble-metal/single-atom contacts, and does not have a large impact
on the nature of the molecule orbitals.

\begin{figure}
\includegraphics[width=0.9\columnwidth]{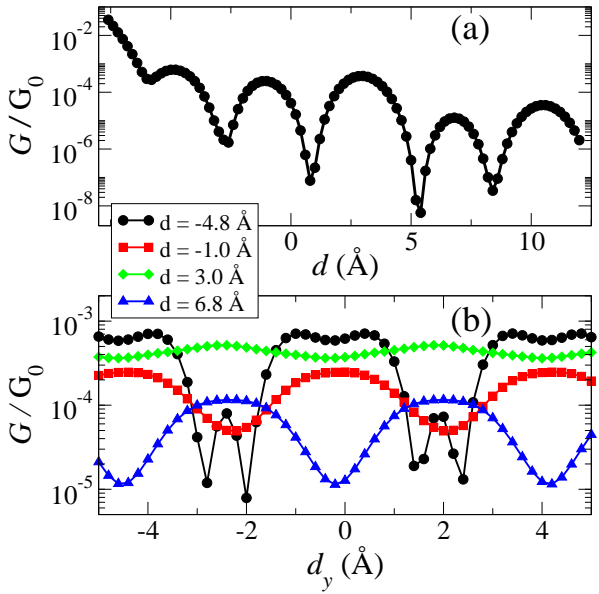}
\caption{\label{Fig4}(Color online) Low voltage conductance $G\,/\,\mathrm{G}_0$ as a
function of changes in the junction. (a) $G/\mathrm{G}_0$ as a function of the gap length $d$;
(b) $G/\mathrm{G}_0$ as a function of the transverse coordinate $d_y$.
$G/\mathrm{G}_0$ is estimated as $T(E_\mathrm{F})$, using only the $\Gamma-$point
in the $k-$summation.}
\end{figure}

The position of the molecular levels of a molecule is not accurately given by conventional
Local Density Approximation-based approaches to DFT by a number of reasons,
including inherent self-interaction errors\cite{Koe08}, an improper description of
quasi-particles\cite{Car12b}, and screening by the metallic electrodes\cite{Nea06}.
As a consequence, the number and position of the true transmission resonances is expected
to be different from those shown in Fig. (\ref{Fig3}). However, the nature and energy
position of the true strongly-correlated quasi-particle states of the molecule depend
only on the hybridization, the net
charge of the molecule and the screening provided by graphene. We note that these do not change
as the molecule is displaced, because of the physisorbed nature of the chemical bond. We deduce that
if we were able to compute the true transmission resonances accurately, then we sould see that
their energy position does not change with molecule position. We conclude that
the universality of the energy
position of the transmission resonance is robust even if strong correlations are included.

\begin{figure}
\includegraphics[width=0.9\columnwidth]{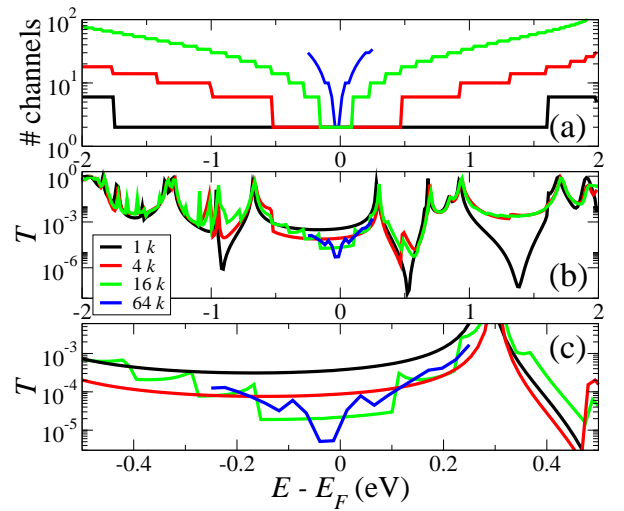}
\caption{\label{Fig5}(Color online) Convergence of the transport
coefficients with the number of $k-$points. (a) Number of channel as a
function of energy; (b) $T(E)$ for the $d=3.0$ \AA\, junction; (c) Blow-up
of $T(E)$ to show the convergency of $T(E_\mathrm{F})$.
Black, red and green lines correspond to 1, 4 and 16 $k-$points, respectively. }
\end{figure}

Fig. (\ref{Fig3}) shows that $T(E)$ computed at the $\Gamma-$point changes by orders of
magnitude at energies $E$ placed between two given BW resonances. A closer inspection
indicates that this is so because Fano-like resonances tend to appear at the gaps between
two BW resonances. In contrast to the BW resonances, the energy position of the Fano-like
resonances shifts as the length of the physical gap increases, or if the molecule is
moved around. Because this also happens at the HOMO/LUMO gap, we expect that the the
$\Gamma-$point estimate of the low-voltage conductance
$G=\mathrm{G}_0\,T(E_\mathrm{F})$ should show
a wide spread. This is confirmed in Fig. (\ref{Fig4}) where $G$ features strong oscillations
as a function of the relative  molecule/graphene position. Notice that these
oscillations  are not strictly periodic as a function of $d$, see Fig. (\ref{Fig4}) (a).
They are however periodic as a function of $d_y$, as shown in Fig. (\ref{Fig2}) (b), with
the period of the graphene lattice along that direction (4.26 \AA) .

We have devised a molecule long enough to bridge the physical electrode gap, and anchor
deep inside the graphene sheets via the planar porphyrine groups. This length requirement
puts stringent conditions on the conjugated nature of the trimer molecule, because
the conductance of a molecule decreases exponentially with its length. Butadiyne chains
are an excellent choice as linking groups because they enhance the molecule's conjugated
nature leading to decay exponents close to zero\cite{victor}. Electrons, however,
can be tranferred between the graphene sheets via several paths. They may hop from
the sheets to the porphyrine units or to the butadiyne molecules. The different
paths lead to destructive interference effects and therefore to Fano-like resonances
in $T(E)$.

We analise now the impact of the $k-$point summation on $T(E)$. Fig. (\ref{Fig5}) (a)
plots the number of channels at the graphene leads as
a function of energy. As the number of $k-$points increases, so does the number of
one-dimensional subbands entering in the energy window. Because the number of channels
decreases steadily as $|E-E_\mathrm{F}|$ decreases, we expect that it is going to be far easier
to converge $T(E)$ at high than at low energies realative to the Fermi level. In other words,
the number of transverse $k-$points needed to obtain accurate estimates of the low voltage
conductance will be very high. Furthermore, as the number of $k-$point tends to infinite, the
number of channels develops a very deep cusp at the Dirac point, as can be inferred from
Fig. (\ref{Fig5}) (a). We expect that this will be transferred to a cusp in
$T(E=E_\mathrm{F})$. To substantiate this reasoning, we show in Fig. (\ref{Fig5})
(b) the evolution of $T(E)$ as the accuracy in the $k-$point summation is improved,
for a given junction arrangement. Notice first that the
energy position of the resonances remains the same, which supports our claim on the
universality of the BW resonances. Second, the Fano-like dips are washed away.
We infer that electron conduction through the new open channels does not suffer destructive
interference effects. Third, the slow accumulation of open channels towards the Fermi
energy implies that the largest loss of accuracy in the computation of $T(E)$ happens
at low energies around $E_\mathrm{F}$. This is specially so if a Fano-like resonance
crosses the Fermi level, leading to a stronger kink at $T(E_\mathrm{F})$.
For instance, the $\Gamma-$point estimates of $T(E_\mathrm{F})$ in Fig.
(\ref{Fig3}) (b) and (\ref{Fig3}) (c)
differ by a factor of 30. Hence, reasonably accurate
estimates of $T(E\sim E_\mathrm{F})$ and of the low-voltage conductance
in graphene-based junctions require the use of a large number of $k-$points,
so that one can infer the asymptotic behaviour.

\begin{figure}
\includegraphics[width=0.9\columnwidth]{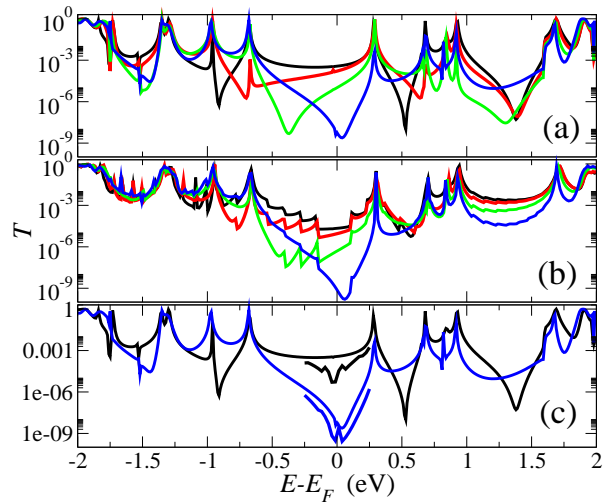}
\caption{\label{Fig6}(Color online) $T(E)$ for $d=$3.0, 3.8, 4.6 and 5.4 \AA\,
for (a) 1 and (b) 16
$k-$points; (c) $T(E)$ for $d=3.0$ and $d=5.4$ \AA\, for 1 and 64 $k-$points are compared.
This graph shows the development of a cusp at $E_\mathrm{F}$.}
\end{figure}

Panels (a) and (b) in Fig. (\ref{Fig6}) show the evolution of the Fano-like resonances as
the molecule moves for 1 and 16 $k-$points, respectively. Clearly,
many resonances do disappear as more channels are added. Notice in particular
how for $d=$4.6 \AA\,
a Fano-like resonance lying in the HOMO/LUMO gap is masked by the conduction through the
$k\ne$0 conduction bands available. The Fano-like resonance for $d=$5.4 \AA\, however,
does not disappear and remains almost unaffected even for large numbers of $k-$ points,
as shown in panel (c). This indicates that only those Fano-like resonances whose dip is very
close to the Dirac point survive when the number of $k-$points increases.
Computing $T(E_\mathrm{F})$ and therefore
the low-bias conductance is complicated by the fact that strictly speaking only the
$k=0$ channel exists asymptotically at $E_\mathrm{F}$ even for a very large
numbers of $k-$points.

In summary, we find that physisorbing molecular wires with large-area
planar anchor groups onto graphene sheets shows universality properties
whereby the molecular BW resonances are placed always at the same energy
position regardless of the relative molecule/sheet position and of the
physical gap length. This universality
property is a consequence of the iso-chemical nature of the physisorbed species.
Our finding is sustained by DFT simulations of several graphene junctions
bridged by monomeric, dimeric and trimeric molecular wires composed of
porphyrine or phthalocyanine molecules. We have used a van der Waals
functional that is suited for the problem at hand. We argue that
even if the energy position of the molecular resonances is fully rearranged
by strong correlations, the new spectra should follow the same universality
behavior. We suggest that the above property could be used to solve the conductance
variability problem inherent to noble-metal-based single-molecule junctions.
This is so because the energy-position of the transmission resonances may be
suitably modified by the application of a gate voltage. The energy spectrum
may also be tailored by synthetic chemistry methods. We stress that, contrary
to the case of noble-metal electrodes, a large number of $k-$points must be
used to find accurate estimates of the low-voltage conductance, which renders
theory predictions a delicate task. The transmission coefficient $T(E)$ develops
a cups at the Dirac point.

We find binding energies
for molecule drift parallel to the physical gap of order 0.1-0.2 eV, which should be enough
to provide mechanical stability to the junction at temperatures below 100 K.
The desired room-temperature mechanical stability of this graphene junctions
seems achievable by attaching side groups, or by increasing the surface area of the anchors.
Furthermore, irregularities at the edges will act as anchoring points.
Longer linker groups than those used in the present study could be synthesized to
better avoid graphene's edges. It would be relevant to design the linker groups so as
to either tailor or avoid the Fano-like resonances.

The research presented here was funded by the Spanish MICINN through
the grants FIS2009-07081 and FIS2012-34858, and by the Marie Curie network NanoCTM.
VMGS thanks the Spanish Ministerio de Ciencia e Innovaci\'on for a
Ram\'on y Cajal fellowship (RYC-2010-06053). JF acknowledges useful
discussions with M. Calame, M. Mayor, M. R. Bryce and C. J. Lambert.
R. Ferrad\'as acknowledges conversations with J. J. de Miguel.


\begin{thebibliography}{References}
\bibitem{Dirtun}
Direct tunneling between electrodes ceases to exist for $d\gtrsim -2$ \AA\, relative
to $d_0$, for our basis choice.
We have taken the trimer molecule away, and checked that the trasmission coefficient $T$
is exactly zero within numerical accuracy for $d\gtrsim -2$ \AA.
\end{thebibliography}
\end{document}